\documentclass[jgrga]{agubook}
\usepackage[dvips]{graphicx}
\setkeys{Gin}{draft=false}
\authorrunninghead{CHEN}
\titlerunninghead{GLOBAL CORONAL WAVES}
\authoraddr{Corresponding author: P. F. Chen (chenpf@nju.edu.cn)}

\begin{document}

\title{Global coronal waves}

\authors{Chen P. F.}

\affil{School of Astronomy and Space Science,
Nanjing University, 163 Xianlin Avenue, Nanjing 210023, China\\
chenpf@nju.edu.cn}

\begin{abstract}
After the {\em Solar and Heliospheric Observatory} ({\em SOHO}) was launched in
1996, the aboard Extreme Ultraviolet Imaging Telescope (EIT) observed a global
coronal wave phenomenon, which was initially named ``EIT wave" after the
telescope. The bright fronts are immediately followed by expanding dimmings. It
has been shown that the brightenings and dimmings are mainly due to plasma 
density increase and depletion, respectively. Such a spectacular phenomenon 
sparked long-lasting interest and debates. The debates were concentrated on two
topics, one is about the driving source, and the other is about the nature of 
this wavelike phenomenon. The controversies are most probably because there may
exist two types of large-scale coronal waves that were not well resolved before
the {\em Solar Dynamics Observatory} ({\em SDO}) was launched: one is a 
piston-driven shock wave straddling over the erupting coronal mass ejection 
(CME), and the other is an apparently propagating front, which may correspond 
to the CME frontal loop. Such a two-wave paradigm was proposed more than 13 
years ago, and now is being recognized by more and more colleagues. In this
paper, we review how various controversies can be resolved in the two-wave
framework and how important it is to have two different names for the two types
of coronal waves. 
\end{abstract}

\begin{article}

\section{Introduction}

\begin{figure}
\noindent\includegraphics[width=38pc]{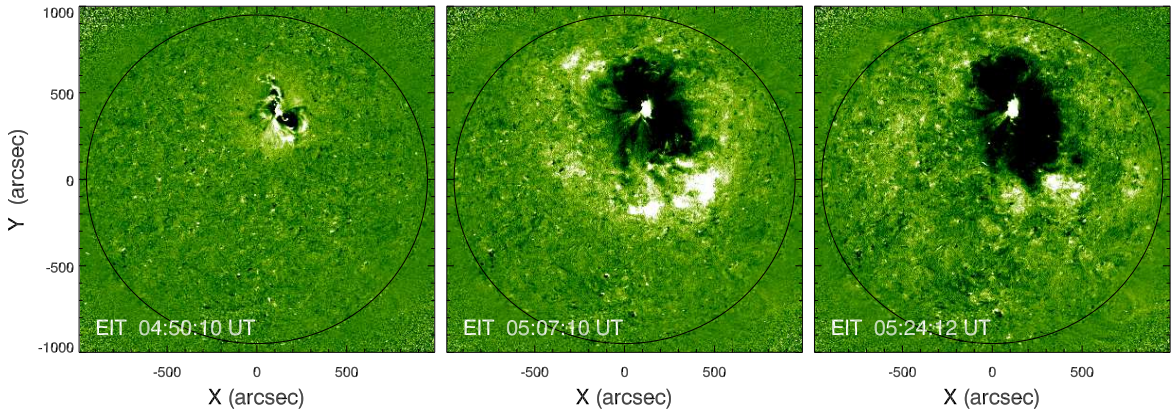}
\caption{Evolution of the 195 \AA\ based difference intensity map observed by
	{\em SOHO}/EIT telescope on 1997 May 12, which was analyzed by
	\citet{thom98}, marking the discovery of ``EIT waves"
	\citep[from][]{chen11}.}
\label{fig1}
\end{figure}

After the launch of the {\em Solar and Heliospheric Observatory} ({\em SOHO}) 
in December 1995, the onboard Extreme Ultraviolet (EUV) Imaging Telescope
\citep[EIT,][]{dela95} was routinely observing the solar corona with a cadence
of 12--15 min at four EUV wavelengths, Fe {\small IX} 171 \AA, Fe {\small XII}
195 \AA, Fe {\small XV} 284 \AA, and He {\small II} 304 \AA. On 1997 May 12, 
the EIT telescope observed an intermediate-scale solar flare in the northern
hemisphere. The flare was registered as C1.3-class according to its peak flux at
1--8 \AA\ measured by the {\em Geostationary Operational Environmental
Satellites} ({\em GOES}). Looking at the original images, one can only identify
a small-sized brightening of the flare and some perturbation in the background,
with nothing special. However, after using a running difference technique,
i.e., to subtract each image with one measured at the previous time so as to
enhance any weak variations, \citet{thom98} discovered an unexpected yet
spectacular phenomenon in the corona. It is seen that a wavelike pattern
propagated outward across the major part of the solar disk, starting from the
source active region of the solar flare, as displayed by Figure \ref{fig1}.
The propagating wave front is immediately followed by an expanding dimming.
The brightening of the fronts is visible at EUV spectral lines with
different formation temperatures, implying that they are mainly due to density
enhancement \citep{will99}. The amplitude of the brightening along the fronts
is averaged to be below 25\% \citep{thom99}. Since the wavelike phenomenon was
discovered with the EIT telescope, it was then called ``EIT wave". Arguing that
it might not be appropriate to name a phenomenon after an instrument, 
colleagues invented several new terms for it \citep[see][for details]{chen12},
such as ``EUV wave" \citep{webb00}, ``global coronal wave" \citep{huds99}, 
``large-scale coronal waves" \citep{ster97}, ``coronal propagating front" 
\citep{schr11}, ``large-scale coronal propagating front" \citep{nitt13}. In
this paper, we tend to use ``EUV waves" for any kind of wavelike phenomena
observed in EUV and ``EIT waves" for the phenomenon discovered by
\citep{thom98}. The coronal EUV waves are different from those waves which are
trapped inside magnetic loops of an active region \citep{asch99, naka99, 
wang15}, confined along large-scale interconnecting loops \citep{liu12}, or in
the coronal holes \citep{bane15}.

For an elastic medium like the solar corona, any local perturbation, e.g.,
in thermal pressure, magnetic field, or velocity, would result in resistance
from the medium. Such resistance is always accompanied by a restoring force, by
which the perturbed part starts to oscillate locally. The oscillation will
perturb the neighboring parts of the medium. Like a chain reaction, a 
perturbation would propagate out as a wave phenomenon, which manifests itself
as a propagating pattern. However, when a propagating pattern is observed, we
sometimes fall into the trap of black-or-white thinking. When we are sure that
the observed moving feature is not mass motion (e.g., when the Doppler velocity
is much smaller than the propagation velocity even after correcting of the
projection effects), we might claim that it is a wave. In this case we forget
the third possibility, i.e., it could be an apparent motion. For example, the 
successive brightening of auroras is a kind of apparent motion, neither a mass 
motion nor a wave. The discovery and the ensuing interpretation of the globally
propagating waves in the solar corona, i.e., ``EIT waves", are an excellent 
example to illustrate how complicated a
propagating pattern would be. This paper is intended to review the progress
on this global coronal wave in the past 18 years. In order not to confuse the
readers, we try to describe the contents in a self-consistent way, which is
inevitably influenced by the author's own opinion. For more balanced 
descriptions among different theoretical models, the readers are referred to
the reviews by \citet{will09}, \citet{warm10}, \citet{gall11}, \citet{zhuk11},
\citet{chen12}, \citet{pats12}, and \citet{liu14}.

\section{Chromospheric Moreton Waves}

It is nearly impossible to talk about ``EIT waves" without mentioning
chromospheric Moreton waves.

In 1960s, with the aim to detect Doppler-shifted features, \citet{more60} used
H$\alpha$-0.5 \AA\ waveband to observe solar flares. As a by product, they
found a bright front in the H$\alpha$-0.5 \AA\ images propagating outward to a
distance of $\sim$10$^5$ km with a velocity of 500--2000 km s$^{-1}$. This 
phenomenon was later widely called ``Moreton waves". The wave fronts are bright
in H$\alpha$ center and the blue wing, while dark in the H$\alpha$ red wing.

These observational features, i.e., the high speed, the long distance, and the
weak amplitude, posed a serious challenge for a straightforward
interpretation, as shown as follows. The solar atmosphere is composed
of a thin photosphere at the solar surface with a thickness of $\sim$500 km, a
slightly thicker chromosphere above the photosphere ($\sim$2000 km), and a vast
corona extending outward to the whole heliosphere. The plasma density drops
drastically from the photosphere to the chromosphere, and further to the corona.
However, the magnetic field decreases with height more gently. As a result, the
Alfv\'en velocity $v_A=B/\sqrt{\mu_0\rho}$ increases rapidly from $\sim$10 km
s$^{-1}$ in the photosphere to $\sim$100 km s$^{-1}$ in the chromosphere, and
further up to $\sim$1000 km s$^{-1}$ in the corona. The H$\alpha$ spectral 
line, with which Moreton waves are observed, is formed in the chromosphere. It
means that Moreton waves are a phenomenon happening in the chromosphere. Because
the Alfv\'en velocity in the chromosphere is of the order of $\sim$100 km 
s$^{-1}$, if Moreton waves, which are observed in the chromosphere with a
propagation velocity of 500--2000 km s$^{-1}$,  are real waves propagating
in the chromosphere, their Alfv\'enic Mach number should be around 5--20, i.e.,
they must be very strong shock waves. However, without continual piston-driving,
such strong shock waves would decay rapidly as they propagate outward so that
they cannot reach a distance of the order of $10^5$ km as found by
\citet{more60}. Therefore, the high velocity and the long propagation distance
of the chromospheric Moreton waves are conflicting characteristics apparently.
Besides, if Moreton waves are such strong shock waves, the amplitude of the wave
should be huge. However, Moreton wave fronts are very faint, and the associated
Doppler velocity is only about 6--10 km s$^{-1}$ \citep{sves76}. These features
puzzled solar physicists for quite a long time. 

\begin{figure}
\noindent\includegraphics[width=38pc]{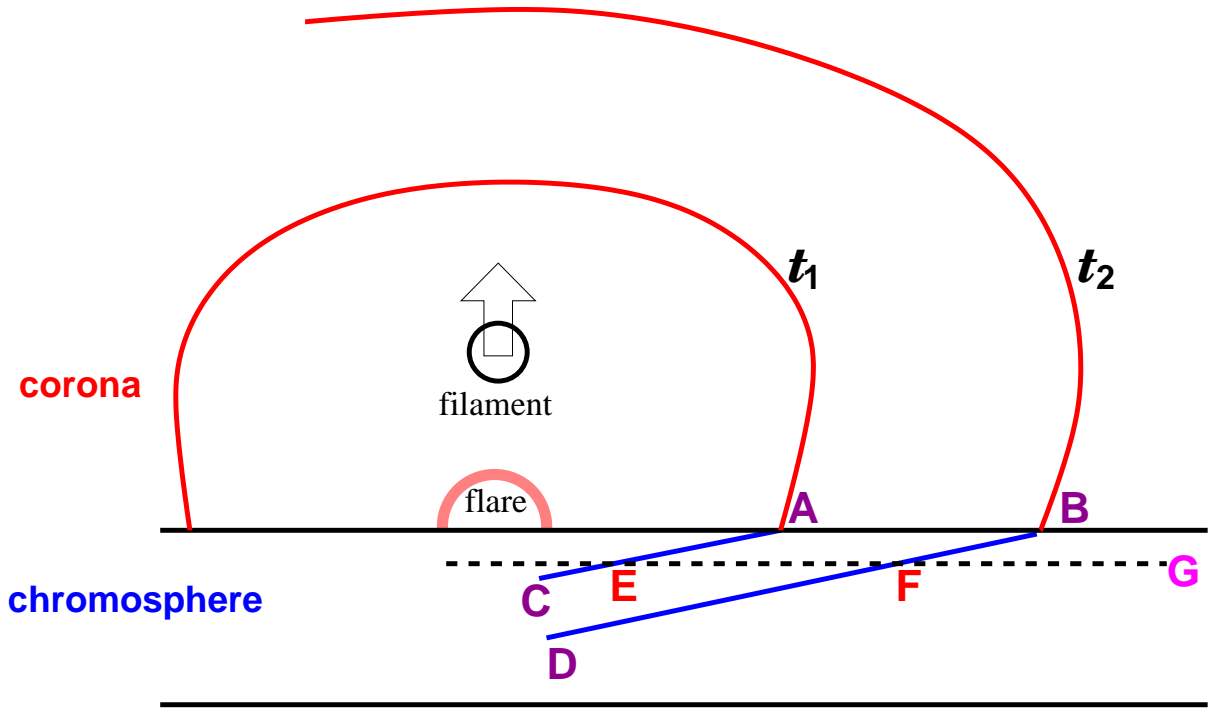}
\caption{Schematic sketch of Uchida's model for H$\alpha$ Moreton waves,
	where the coronal fast-mode MHD wave (or shock wave) sweeps the
	chromosphere, generating an apparent propagation of the H$\alpha$
	Moreton wave from points E to F with the same speed as the coronal
	fast-mode wave. It is noted that since the amplitude of the
        fast-mode MHD wave front in the chromosphere decreases drastically
	as the wave penetrates deeper, not every part of the wave front
	({\it blue lines}) contributes to the observed Moreton wave.}
\label{fig2}
\end{figure}

Several possibilities were put forward, but the basic idea was the same: Moreton
waves are related to waves propagating in the corona \citep{carm64, meye68}. A
proper treatment was conducted by \citet{uchi68}, who proposed that the pressure
pulse in the solar flare generates a fast-mode magnetohydrodynamic (MHD) wave
propagating in the corona. The skirt of the wave front sweeps the chromosphere,
pushing the chromospheric material downward, by which a Moreton wave front is
formed. Such a process is illustrated in Figure \ref{fig2}. The basic framework
of Uchida's model is correct, the main modification we made in Figure 
\ref{fig2} is that the coronal counterpart of a Moreton wave is driven by the 
eruption of a flux rope with a filament at the bottom (note that the filament 
is often identified as the core of a CME), i.e., the coronal counterpart of any
Moreton wave is a piston-driven shock wave straddling over the ejecta, rather 
than a blast wave initiated by the pressure pulse in the accompanied solar 
flare \citep[see also][]{cliv99, chen02}.

According to the modified Uchida model in Figure \ref{fig2}, the erupting flux
rope drives a fast-mode coronal shock wave ({\it red lines}) propagating 
outward. During the time interval from $t_1$ to $t_2$, the footpoint of the
coronal shock wave moves from point A to point B with a velocity of $\sim$1000 
km s$^{-1}$.  At the same time, the wave front in the chromosphere also 
propagates from CA to DB with a velocity of $\sim$100 km s$^{-1}$. Since 
the fast-mode wave velocity in the chromosphere is $\sim$10 times smaller than
that of the coronal fast-mode wave, the propagation
distance between the two fronts DB and CA would be $\sim$10 times shorter than
that of the two coronal fronts at the time $t_2$ and $t_1$. As a result, the 
real wave fronts in the chromosphere would be strongly oblique as shown by the
blue lines in Figure \ref{fig2}. The forward-inclined wave front was inferred by
observations \citep{vrsn02, gilb08}. According to the radiative transfer 
calculations, the emissions of the H$\alpha$ line profile at different
wavelengths come from different layers of the chromosphere, say, the H$\alpha$
line center is formed in the upper chromosphere, whereas the H$\alpha$ line
wings are formed in the lower chromosphere \citep{val81}. In observations,
Moreton waves are generally detected with H$\alpha$ filtergrams using a narrow
waveband in the H$\alpha$ profile, e.g., H$\alpha$-0.5 \AA\ \citep{more60}. The
emission is mainly from a certain layer of the chromosphere. Without loss of
generality, we can assume the chosen H$\alpha$ waveband is formed at the
level G, which intersects with the two chromospheric fronts at points E and F
as indicated by the dashed line in Figure \ref{fig2}. As a result, an H$\alpha$
Moreton wave front would be seen to propagate from point E to point F, which is 
equivalent to the distance from point A to point B, i.e., the H$\alpha$ Moreton
wave has the same velocity as the footprint of the coronal fast-mode wave, though
lagging behind. Such a spatial delay of about tens of Mm was clearly found in
observations by \citet{vrsn02} and recently by \citet{whit13}, and was reproduced
by simulations as well \citep{chen05a}. The co-spatiality between the H$\alpha$
Moreton wave and the coronal shock wave mentioned in some observations 
\citep[e.g.,][]{asai12, shen12} is just an approximation due to limited spatial
resolution. It
is also seen that the wave fronts in the chromosphere are nearly horizontal due
to the significant difference between the fast-mode MHD wave speeds in the 
corona and in the chromosphere. Hence it is expected that the Doppler velocity 
of the chromospheric plasma in the downstream of the fast-mode MHD wave, which
is mainly perpendicular to the wave front, is downward.
The downward motion results in a red shift of the H$\alpha$ line profile,
producing brightening in the blue wing of the H$\alpha$ line and
darkening in the red wing. The intensity at H$\alpha$ line center is also
slightly increased. It was shown that the intensity variation due to the red
shift of the chromosphere peaks at H$\alpha\pm 0.45$ \AA\ \citep{chen05a},
meaning that H$\alpha\pm 0.45$ \AA\ might be the best passband for the
detection of Moreton waves using filtergrams. The essence of Uchida's model is
that the chromospheric Moreton wave itself is not a wave, it is the footprint
of the fast-mode coronal wave. Since the fast-mode wave speed in the corona is
of the order of 1000 km s$^{-1}$, when an H$\alpha$ Moreton wave is 
observed to travel with a speed higher than 1000 km s$^{-1}$, the 
coronal counterpart of the Moreton wave is either a simple wave or a weak shock
wave. This then explains why it can propagate for a long distance and the
amplitude of the wave front is small.

Based on Uchida's model, a fast-mode coronal Moreton wave is expected to exist,
which was called coronal Moreton wave by \citet{kruc99}, \citet{tors99}, and
\citet{thom99}.

\section{Are ``EIT waves" the expected coronal Moreton waves?}
 
When coronal ``EIT waves" were discovered \citep{mose97,thom98}, they were
thought to be the long-awaited coronal Moreton waves, i.e., fast-mode MHD waves
or shock waves \citep{thom99,wang00,wu01,ofma02,selw13}. There are several
reasons for the deduction. First, any coronal mass ejection (CME)/flare 
eruption should excite fast-mode MHD waves, and only ``EIT waves" were 
observed, therefore, ``EIT waves" {\it should} be fast-mode MHD waves.
Second, at one moment in two events, the ``EIT wave" front was found to be
nearly cospatial with the H$\alpha$ Moreton wave front \citep{thom00,pohj01}. 
However, the observed ``EIT waves" showed more features that cannot be 
accounted for in terms of fast-mode MHD waves.

If ``EIT waves" were the fast-mode coronal Moreton waves, there should be strong
correlation between the ``EIT wave" speed and the speed of type II radio bursts
since the latter result from the same shock wave as the Moreton wave
\citep{uchi74}. However, \citet{klas00} found that
there is a lack of correlation between the ``EIT wave" speed and the speed of 
type II radio bursts. Furthermore, \citet{klas00} and \citet{zhan11} revealed 
that the ``EIT wave" speed is typically three times smaller than the Moreton 
wave speed (or the propagation speed of the type II radio burst). More
seriously, ``EIT wave" speed can be well below the sound speed in the corona 
\citep{trip07, thom09}, sometimes even down to $\sim$10 km s$^{-1}$
\citep{zhuk09}.

If ``EIT waves" were the fast-mode coronal Moreton waves, it is then expected
to see a strong positive correlation between the ``EIT wave" speed and the
local magnetic field strength if the plasma density does not change much.
However, \citet{yang10} found that the ``EIT wave" speed is negatively 
correlated with the magnetic field strength. It is noticed that \citet{zhao11}
obtained the opposite conclusion by checking the correlation between the ``EIT 
wave" speed and the local fast-mode wave speed ($v_{\rm f}$). On the one hand, 
it is more appropriate to compare the ``EIT wave" speed with $v_{\rm f}$ as in
\citet{zhao11} rather than just the magnetic field strength as done by
\citet{yang10}. On the other hand, the ``EIT wave" event analyzed by
\citet{zhao11} was on the east limb, and the photospheric magnetogram used in
their modeling was actually measured about one week later, hence it might not
represent the real situation when the ``EIT wave" was propagating. Besides, 
the assumed coronal density in \citet{zhao11} is so high and the resulting 
Alfv\'en speed is so low, e.g., even $<$100 km s$^{-1}$, that the ``EIT wave" in
their model would be a strong shock wave, which might not be able to explain 
the small amplitude of the ``EIT wave" fronts.

If ``EIT waves" were the fast-mode coronal Moreton waves, they would be
concentrated toward regions of smaller fast-mode speeds (probably weaker
magnetic fields, assuming that the plasma density does not vary much) due to
refraction, showing no systematic pattern related to the helicity of the 
source active region. However, with 8 examples {\it Attrill et al.} [\citeyear{attr07}; \citeyear{attr14}] revealed
that the rotation of ``EIT wave" fronts, a feature discovered by 
\citet{podl05}, shows a systematic helicity rule, i.e., being anticlockwise 
when the source active region has negative helicity and clockwise when the 
source active region has positive helicity. The sense of rotation is the same 
as the erupting filaments, which are untwisting during eruptions \citep{chen11}.

There are many other features of ``EIT waves" that cannot be explained simply by
the fast-mode wave model. For example, \citet{chenf11} found that the plasma
outflow in a small coronal hole diminished after an ``EIT wave" passed by.
\citet{chen09b} and \citet{dai10} claimed that the ``EIT wave" front is
cospatial with the CME frontal loop. Well before these work, it is 
\citet{dela99} and \citet{dela00} who first challenged the fast-mode wave model
for ``EIT waves". In three events, they found that a stationary wave front was
located at a magnetic separatrix, across which the magnetic field strength 
might change smoothly but the magnetic fields on its two sides belong to
different magnetic systems. This feature is difficult to be accounted for by
the fast-mode wave model. They explained it to be due to the opening of the
closed magnetic field lines.  
It is noted in passing here that the stationary EUV front mentioned above might
not be concluding evidence against the fast-mode MHD wave. For example, 
\citet{kwon13} found that two stationary EUV fronts located on the two
footpoints of a streamer brightened soon after a fast-mode MHD wave swept
the upper streamer. They explained the stationary EUV fronts as trapped 
fast-mode waves due to the deflection of the streamer. On the other hand,
it is commented by B. Vr{\v s}nak (2015, private communication) that these two
fronts are actually moving away from each other slightly, rather than being 
stationary, and the two separating fronts might be similar to flare ribbons 
due to magnetic reconnection of the current sheet above the streamer, which
was triggered by the transiting fast-mode wave.

\section{Toward a better model}\label{model}

\begin{figure}
\noindent\includegraphics[width=38pc]{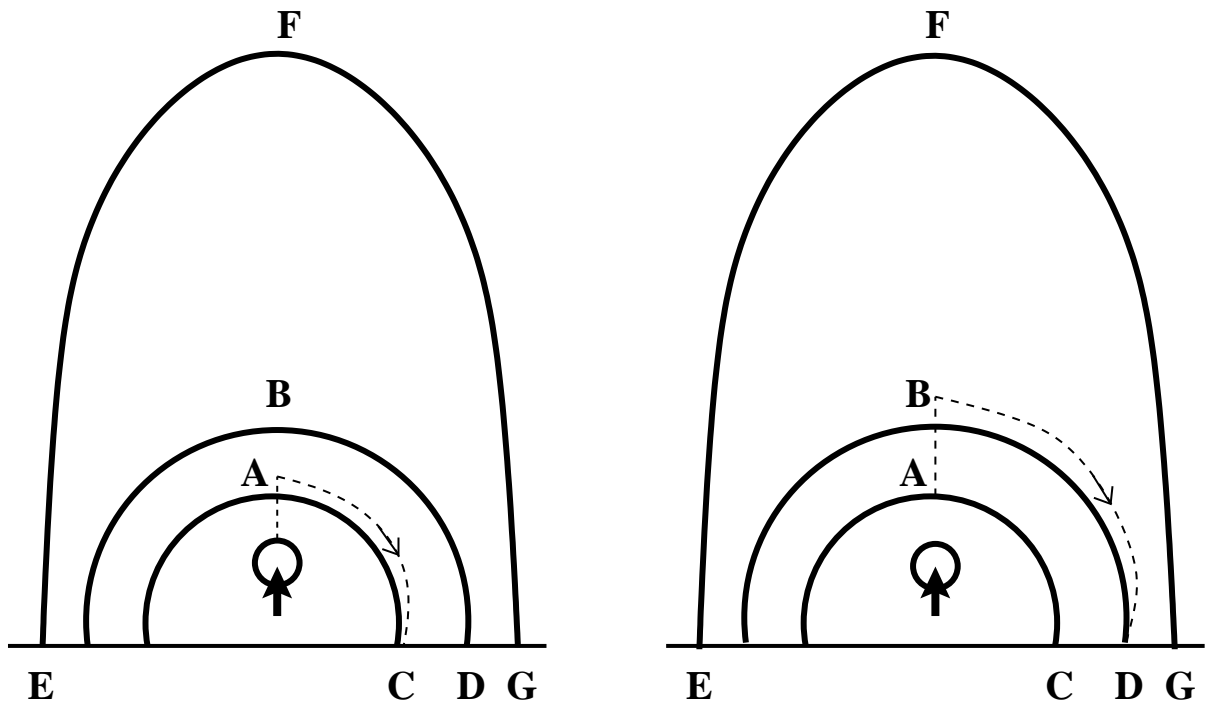}
\caption{Schematic diagram illustrating how the magnetic fieldline stretching
pushed by the erupting flux rope is transferred from the top to the footpoint
of each field line so that ``EIT wave" fronts are formed successively, from
point C to point D at two different times
	\citep[from][]{chen05b}.}
\label{fig3}
\end{figure}

Inspired by the questioning by \citet{dela00}, \citet{chen02} performed MHD
numerical simulations to examine what wave phenomena would appear during CME
eruptions. In their MHD numerical results, as a flux rope erupts, two types of 
wavelike phenomena appear in the corona. The faster one is a piston-driven 
shock wave straddling over the erupting flux rope. The leg of this wave travels
outward at a speed 773 km s$^{-1}$ in the horizontal direction, and was 
proposed to be the coronal Moreton wave. The slower one also straddles over 
the erupting flux rope, but behind the coronal Moreton wave. The horizontal
speed of its wave front is only 250 km s$^{-1}$. \citet{chen02} proposed that
the slower wave 
corresponds to the ``EIT wave" observed by \citet{thom98}, which is about 3 
times slower than the coronal Moreton wave, consistent with observations
\citep{klas00, zhan11}. In order to explain the unexpected ``EIT wave", 
\citet{chen02} noticed that as a flux rope erupts, all the closed magnetic 
field lines straddling over the flux rope would finally be stretched out, 
successively from the inner field lines to the outer field lines. For any 
individual field line, the stretching starts from the top, and is then 
transferred down to the footpoint of the field line. Based on this intuition, 
\citet{chen02} proposed the magnetic fieldline stretching model for ``EIT 
waves", which is illustrated by Figure \ref{fig3}. Note that density
enhancement is formed outside the field line when any part of the field lines
is newly stretched, so all the stretching patches at one
moment form a pattern, which is a domelike structure straddling over the 
erupting flux rope \citep{chen09b}, as found from observations \citep{vero10}.

According to this idea as illustrated in Figure \ref{fig3}, the corresponding
apparent velocity of the ``EIT wave"
is $v_{\rm EIT}=CD/\Delta t$, where the time difference between the successive
formation of the ``EIT wave" fronts at points D and C is expressed as
$\Delta t=AB/v_{\rm f}+(BD-AC)/v_{\rm A}$, where $v_{\rm f}= \sqrt{v_{\rm A}^2
+c_{\rm s}^2}$ and $v_{\rm A}$ are the fast-mode wave speeds
across and along the field lines, respectively, $v_{\rm A}$ is the Alfv\'en
speed, and $c_{\rm s}$ is the sound speed. For any prescribed magnetic 
distribution, we can always quantitatively derive the corresponding ``EIT wave"
velocity. For simplicity, \citet{chen02} assumed an ideal case where all the
magnetic field lines overlying the flux rope are concentric semi-circles, and the
corresponding ``EIT wave" velocity is found to be $v_{\rm EIT}=0.34v_{\rm f}$,
i.e., the ``EIT wave" speed is roughly 3 times smaller than the local fast-mode
wave speed. This is consistent with the observations \citep{klas00, zhan11}, as
well as the MHD numerical simulations presented in \citet{chen02}. The large Doppler
velocity and spectral line width immediately behind the ``EIT wave" in the
numerical simulations are consistent with those observed by \citet{chenf10}.
Later, \citet{chen05b} illustrated how the magnetic fieldline stretching model 
can explain why ``EIT waves" stop near magnetic separatrices and why there is 
no correlation between the ``EIT wave" velocity and the speed of the type II 
radio bursts. One important prediction of the magnetic fieldline stretching
model is that, given a sufficiently high time cadence, any EUV imaging
telescope can observe two wavelike patterns, one is faster and the other is
slower. It should be noted that the 3-fold relation between the faster wave
and the slower wave is valid when the magnetic field lines overlying the flux 
rope are concentric semi-circles. If the magnetic field lines are elongated in 
the upward direction, the velocity ratio between the two waves is not 
necessarily to be 3. Therefore, the velocity ratio of 4 between two EUV waves
found by \citet{zhen12a} does not conflict with the magnetic fieldline 
stretching model. It should be pointed out that in Figure \ref{fig3} we
assume for simplicity that the stretching is transferred from point A to C
for the first EIT wave front, and then from point A to D via B for the next
EIT wave front. In reality,
the transfer of the stretching from point A to D may detour around point B,
which would result in the widening of the EIT wave front during propagation
as demonstrated by \cite{chensh02}. Furthermore, it was proposed that a CME
may experience a short period of overexpansion in the low corona 
\citep{pats10}. In such a spherical piston-driven case, the stretching is
transferred directly in the horizontal direction, and the EIT wave velocity
would be identical to the fast-mode wave speed. This may explain why the 
slow ``EIT wave" and the fast-mode wave are coupled in the early stage
\citep{cheng12}.

Since then, many other models have been proposed to explain the slow ``EIT 
waves". For example, \citet{attr07}
suggested that EIT wave fronts are produced when the erupting magnetic field
lines reconnect with a series of low-lying antiparallel magnetic loops, as
depicted by Figure 4 in their paper. Our opinions on this model are: (1) ``EIT 
waves" are not a low-lying phenomenon. They often extend $\sim$90 Mm above
the solar surface \citep{pats09b}, and sometimes are observed to have a 
domelike structure \citep{vero10}. This issue has not yet be addressed in the 
successive reconnection model. (2) Whereas it is hard to see that all
low-lying magnetic loops are antiparallel with the erupting field lines, there
are occasionally antiparallel magnetic loops in the background, in which case
magnetic reconnection between the erupting magnetic field and the background
magnetic field can happen, as demonstrated by \citet{cohe09}. Therefore, the
combination of this successive magnetic reconnection model and the magnetic
fieldline stretching model can account for many observational features of ``EIT 
waves". \citet{dela08} proposed a current shell model for ``EIT waves". In this
model, ``EIT waves" correspond to the current shell between the erupting core
magnetic field and the potential background magnetic field. One feature of this
model is that the footpoints of the current shell are fixed, therefore, the
authors associate the propagating ``EIT wave" fronts with the current shell at
a certain height. \citet{down12} reproduced such a current shell structure,
which is confined to the region of CME itself and mismatches the outer EUV
wave front. Our intuition is that the current-shell model might be applied to
explain some looplike structures near the source active region during CMEs.
\citet{will07} proposed an idea related to slow-mode soliton waves, with the 
aim to explain the coherence of the propagating ``EIT wave" fronts. One 
difficulty is that ``EIT waves" propagate across magnetic field lines, whereas 
slow-mode waves propagate along magnetic field lines. Based on MHD numerical 
simulations, \citet{wang09} proposed that ``EIT waves" might be due to 
the joint impact of the slow-mode shock waves and the vortices behind the
erupting flux rope.

\section{Evidence of two types of EUV waves}

\begin{figure}
\noindent\includegraphics[width=38pc]{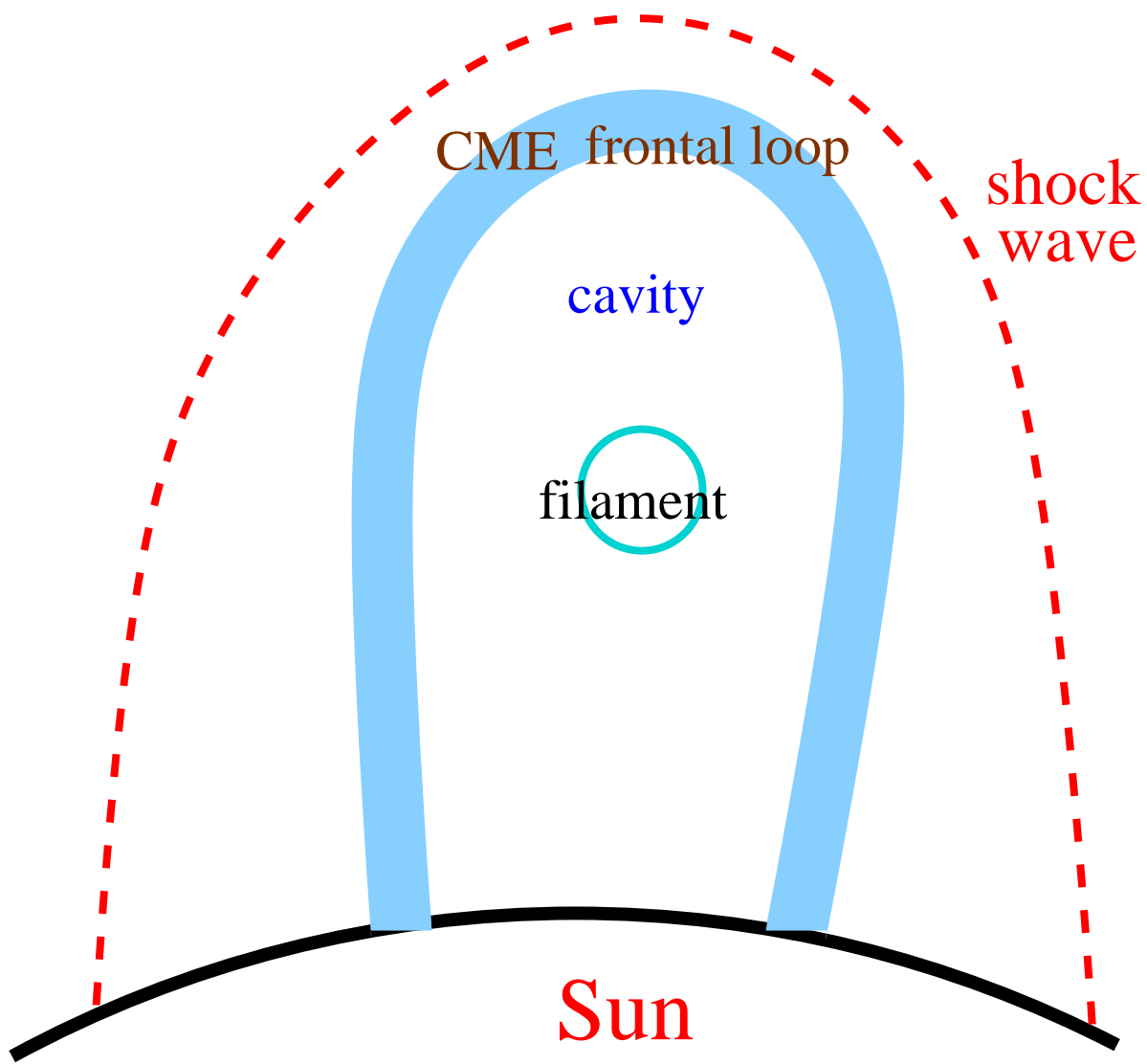}
\caption{A sketch of the classical 3-component structure (a frontal loop, a
	bright core representing an erupting filament, and the cavity in
	between) of a CME observed in white-light.}
\label{fig4}
\end{figure}

It is widely accepted that a full-fledged CME is composed of three components,
i.e., a convex-outward frontal loop, an embedded bright core representing the
erupting filament, and a cavity in between \citep{forb00}. Besides, a 
piston-driven shock, which is a fast-mode shock wave, frequently straddles over
the frontal loop, producing type II radio bursts \citep{wild63} and manifesting
itself in white-light coronagraph images \citep{vour03}. All these observational
results have led to a paradigm for CMEs as depicted by Figure \ref{fig4}, though
one or two parts in the schematic sketch might not be present in an
individual event. Now suppose such a complex structure as depicted in Figure 
\ref{fig4} is observed from above by EUV telescopes, what wavelike patterns can
be seen? We would expect to see both the piston-driven shock wave and the CME
frontal loop propagating outward \citep{pats12}. That is to say, even with an
imaginary test, we would expect to see two wavelike phenomena in the EUV images
of the solar disk based on the widely accepted CME model, with the first wave
being faster than the second one.

From the observational point of view, the evidence of two EUV
waves already existed in the {\it SOHO}/EIT observations, despite the low
cadence of the telescope \citep{warm11}. For example, \citet{bies02} found a
special subclass of ``EIT waves", whose fronts are very sharp, in contrary to 
the diffuse fronts of ordinary ``EIT waves". This subclass of ``EIT waves" 
should be the coronal Moreton waves, and indeed, in two events, these sharp
``EIT waves" were found to be nearly cospatial with H$\alpha$ Moreton waves at a
single moment \citep{thom00, pohj01}. Note here again that the cospatiality 
occurs at only one moment, since H$\alpha$ Moreton waves are usually visible
for $<$10 min \citep{fran13}, whereas the time cadence of the EIT telescope is
12--15 min. Unfortunately, the occasional cospatiality between the sharp ``EIT 
wave" front and the H$\alpha$ Moreton wave can be easily considered to be the 
strong evidence to support the fast-mode wave model for ``EIT waves". To explain
the velocity discrepancy between Moreton waves and ``EIT waves" in the 
fast-mode wave model, it was then
argued that the fast-mode wave decelerates later in the quiet Sun region
\citep{wu01}. However, both the H$\alpha$ \citep{eto02} and radio 
\citep{whit05} observations did not show any significant deceleration of the 
Moreton wave.

The first observational evidence of two EUV waves was discovered by 
\citet{harr03}, who used the high-cadence (1--2 min) data from the {\it
Transition Region and Coronal Explorer} ({\it TRACE}) satellite. They found a
brighter wave front moving with a velocity of $\sim$200 km s$^{-1}$ and a
weaker wave front moving with a velocity of $\sim$500 km s$^{-1}$. It is a
pity that the authors did not use a time-distance plot to clearly display the
separation of the two waves so that their statement was questioned later. For
example, \citet{will06} analyzed the same event and argued that only one wave
exists. She considered the faster weak front ahead as a Gaussian
extension of the slower bright front behind. However, in terms of 
a simple wave with certain extension as mentioned by \citet{will06}, the wave 
portions with larger amplitudes should travel faster than those with smaller 
amplitudes. In this sense, the brighter and fainter fronts recognized by both 
\citet{harr03} and \citet{will06} are more likely to be two separate fronts.

After the {\it Solar Terrestrial Relations Observatory} ({\it STEREO}) twin
satellites were launched in 2006, its onboard Extreme Ultra-Violet Imager 
(EUVI) with a higher cadence (2.5 min) did not bring any breakthrough in
resolving the theoretically predicted two types of EUV waves. Instead, most 
papers based on {\it STEREO}/EUVI data tend to support the
fast-mode wave model for ``EIT waves" \citep{long08,gopal09}. For example,
\citet{pats09a} analyzed the first quadrature observations of an EIT wave event,
and concluded that EIT waves are fast-mode waves. The reason is that they 
found that the EIT wave front is outside the CME bubble. We tend to think that
their EIT wave front is cospatial with the CME bubble, which is especially clear
at 05:55 UT in their Figure 1. The difficulty in matching an EIT wave with
the CME bubble lies in the lack of common field of view between the EUV
imager and the coronagraph.

\begin{figure}
\noindent\includegraphics[width=38pc]{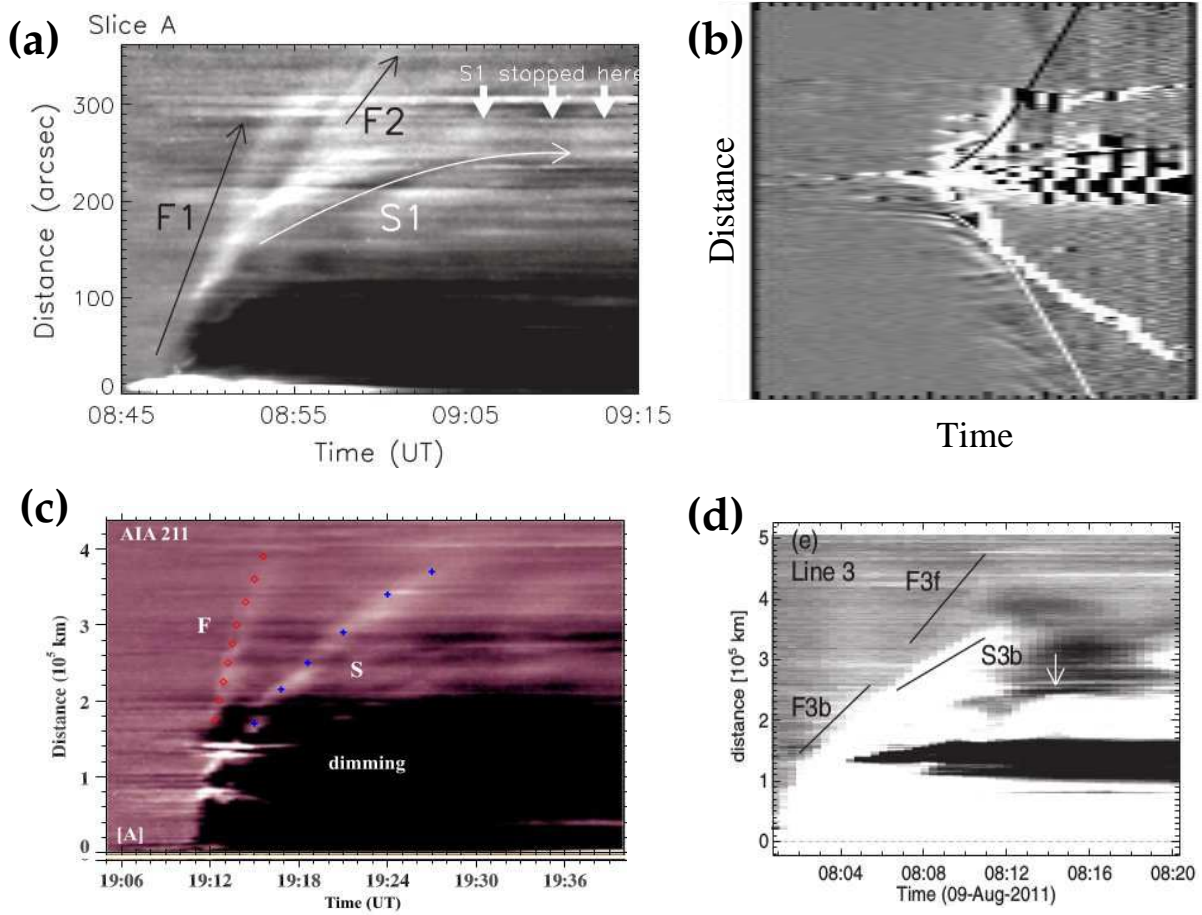}
\caption{A collection of several time-distance diagrams of the EUV intensity
	showing two types of EUV waves propagating with different velocities.
	(a) from \citet{chenwu11}; (b) adapted from \citet{schr11};
	(c) adapted from \citet{kuma13}; and (d) from \citet{asai12}.}
\label{fig5}
\end{figure}

\begin{figure}
\noindent\includegraphics[width=38pc]{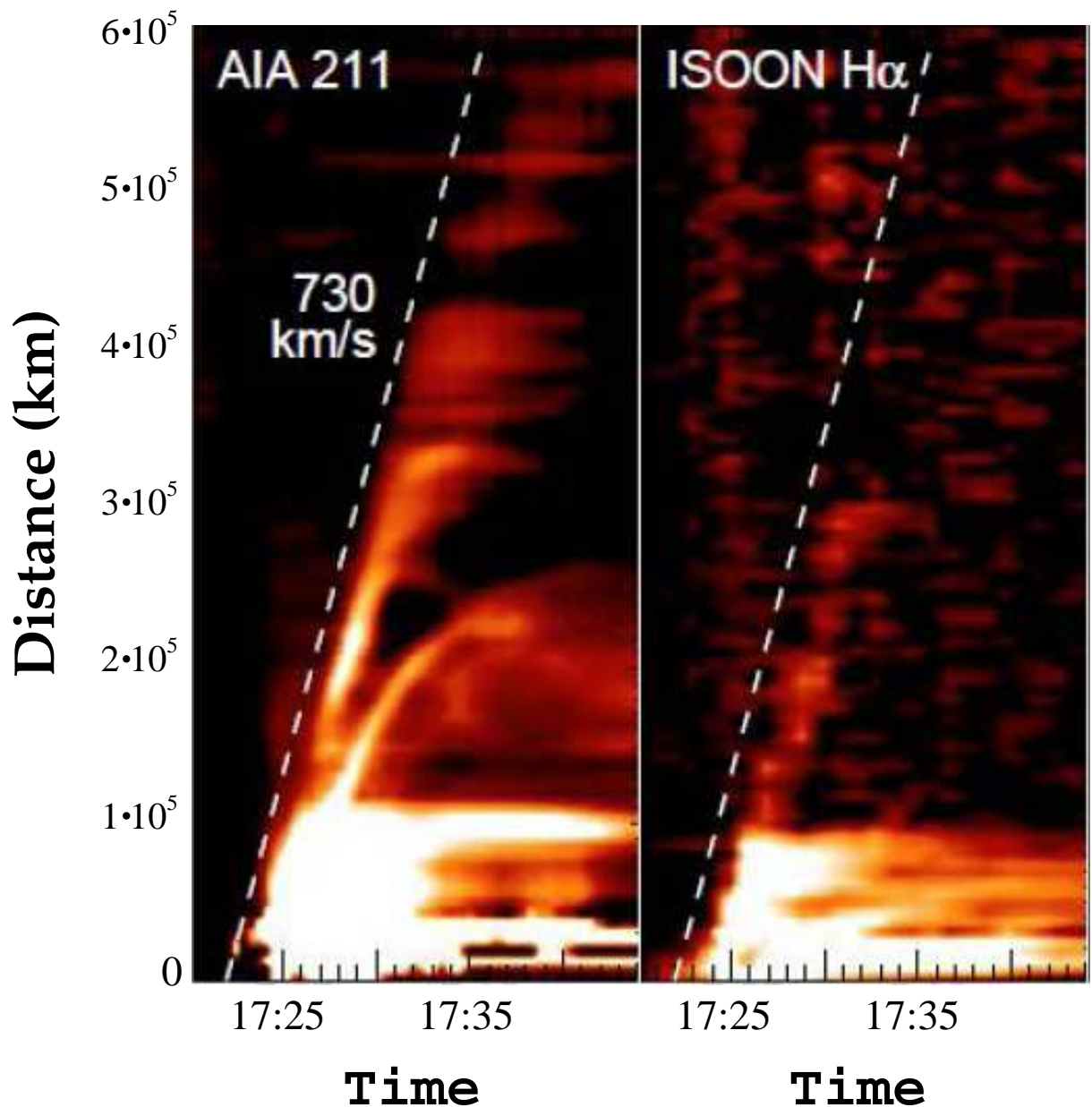}
\caption{Time-distance diagrams of the AIA 212 \AA\ intensity ({\it left}) and 
	the {\it ISOON} H$\alpha$ intensity ({\it right}), clearly
	demonstrating the faster EUV wave is nearly cospatial with the 
	H$\alpha$ Moreton wave and the slower EUV wave is much behind 
	\citep[adapted from][]{whit13}.}
\label{fig6}
\end{figure}

The real breakthrough on this topic came after the launch of the {\em Solar
Dynamics Observatory} ({\em SDO}) satellite in 2010. The unprecedentedly high
cadence ($\sim$12 seconds) and high spatial resolution of the aboard 
Atmospheric Imaging Assembly (AIA) telescope unveiled many details of the
kinematics of various EUV waves. Compared to previous instruments, the 
spatiotemporal resolution of the {\em SDO}/AIA telescopes is improved so
much that the new observations revealed too many subtle features that are
sometimes difficult to interpret straightforward \citep{liu10}. However, in a
surprisingly weak CME/flare event, the long-awaited two types of waves were
finally disclosed clearly \citep{chenwu11}. The time-distance plot of the EUV
intensity of this event is displayed in panel (a) of Figure \ref{fig5}, where
the faster wave has a velocity of 560 km s$^{-1}$, whereas the slower wave has
a velocity of 190 km s$^{-1}$. The authors argued that the faster wave
corresponds to the coronal Moreton wave, therefore, it is of fast mode; the 
slower one corresponds to the diffuse ``EIT wave" as discovered by 
\citet{thom98}. Since then, more and more events were revealed to have a
faster wave ahead of a slower wave, which is best visible from time-distance
plots \citep{schr11, asai12, cheng12, shen12, kuma13, shen13}. Some of these 
results are displayed in panels (b--d) of Figure \ref{fig5}. In particular,
\citet{asai12} caught the rare chance to observe EUV waves and an H$\alpha$
Moreton wave simultaneously (panel d in Figure \ref{fig5}), and they confirmed
that the faster EUV wave is nearly cospatial with the H$\alpha$ Moreton wave, whereas
the slower wave is behind. An even better example is the 2011 February 14
event analyzed by \citet{whit13}. Figure \ref{fig6} illustrates the
time-distance plots in the AIA 211 \AA\ and the {\it ISOON} H$\alpha$
wavebands. Their result convincingly confirmed that H$\alpha$ Moreton wave
corresponds to a faster EUV wave, which is followed by a slower yet brighter
EUV wave. The two-wave paradigm was recently supported by 
3-dimensional MHD simulations \citep{down12}. In January 2014, a group of
``EIT wave" colleagues were gathering in Bern for an International Space 
Science Institute (ISSI) workshop led by D. Long and S. Bloomfield. Most
participants agree that there are two (or at least two) types of EUV waves.

\section{What caused the confusion?}

The nature of ``EIT waves" has been debated for $\sim$16 years. It is time to
comb the complications and think about what factors led to the controversy
among the community.

With respect to H$\alpha$ Moreton waves, the serious problem is the
rarity of the events. Even during only 15 months from 1997 March 24 to 1998 June,
about 176 EUV waves were detected by {\it SOHO}/EIT \citep{thom09}. On the
contrary, only dozens of H$\alpha$ Moreton waves have been observed during
the past 55 years. Therefore, it is very rare to find simultaneous Moreton
wave and EUV wave observations. The reasons for the rarity of Moreton wave
events include: (1) Only when the coronal shock wave is strong enough, its
skirt can heavily sweep the chromosphere so as to generate significant 
downward motions of the dense chromosphere, i.e., perturbations should exist
in many CME/flare events, but in most cases they are not strong enough to
produce Doppler shifts of the H$\alpha$ line; (2) Moreton waves were 
discovered in H$\alpha$-0.5 \AA\ \citep{more60}, and it was calculated that
the passband around H$\alpha\pm 0.45$ \AA\ is the best suitable for detecting
Moreton waves. However, most of the current H$\alpha$ telescopes are using 
H$\alpha$ line center, with only a few exceptions \citep[e.g.,][]{ueno04,
fang13}.

With respect to EUV waves, the serious problem is the observing cadence.
In the {\em SOHO}/EIT era, i.e., from 1996 to 2006, the cadence of the EUV
imaging observations was too low ($\sim$15 min). However, H$\alpha$ Moreton 
waves have relatively short lifetimes, which are generally less than 10 min 
\citep{will07}. The 12-min lifetime of the Moreton wave analyzed by 
\citet{bala07} is exceptionally long. Therefore, none or at most one front of
the Moreton wave can be captured by the {\em SOHO}/EIT telescope. What the 
{\em SOHO}/EIT telescope observed was mainly the slower type of EUV waves. In
our opinion, most of them are the EUV counterpart of the CME frontal loop, 
rather than fast-mode waves or shock waves.

In the {\em STEREO}/EUVI era, i.e., from 2006 to 2010, the observational cadence
was increased to 2.5 min. In principle, several fronts of an Moreton wave event
at successive times can be captured, in addition to the slower wave. However,
it was demonstrated by \citet{chenwu11} that, at least for the EUV wave event
on 2010 July 27, only when the observational cadence is
less than $\sim$70 seconds can the two EUV waves be clearly distinguished. This
is because both waves originate from the source active region, and they are
still not far enough from each other during the lifetime of the faster EUV wave.
Only occasionally, for example, when the quiet Sun magnetic field is very weak
and the fast-mode wave speed is not very large compared to the sound speed, the
two EUV waves might be discernable \citep{grec11,down12}.

During the current {\em SDO}/AIA era, i.e., after 2010, it has been demonstrated
that two types of EUV waves can be distinguished well. We were anticipating that
a converging consensus should be reached soon \citep{chen12}. However, the
current state of the ``EIT wave" research is still confusing. In our 
understanding, the confusion in the {\em SDO}/AIA era is due to the
terminology, i.e., ``EUV waves" or ``coronal waves" were used for both wave
phenomena indiscriminately. Among case studies, some events correspond to the
fast-mode MHD waves or shock waves \citep[e.g.,][]{li12, yang13}, whereas some 
others correspond to the diffusive ``EIT waves" \citep[e.g.,][]{zhen12b}. 
Similarly, among statistical studies, some samples mostly consist of the
fast-mode MHD waves or shock waves, probably mixed with a few slower wave
events. For example, \citet{nitt13} focused on the fastest wave pattern
in the time-distance plot of each event observed by {\em SDO}/AIA, which they
call large-scale coronal propagating fronts (LCPFs). That is why the mean 
velocity of their LCPFs is as high as 644 km s$^{-1}$. However, the EUV waves
in some other studies mainly belong to the slower type of ``EUV waves", e.g.,
\citet{muhr14} investigated the EUV wave kinematics using the {\em STEREO}/EUVI
data and found that the arithmetic mean of the linear velocity is only $254\pm 
74$ km s$^{-1}$. Some colleagues might think that the statistical results of
\citet{muhr14} and \citet{nitt13} are contradictory.  In our personal 
view, the apparent discrepancy immediately disappears if we consider the waves
in \citet{muhr14} as the slower component and the LCPFs in
\citet{nitt13} as the faster component of the two-wave paradigm. In this sense,
the waves called ``coronal propagating fronts" by \citet{schr11} belong to the 
slower component, very different from the LCPFs in \citet{nitt13}. Note that, 
in our viewpoint, a few events in \citet{nitt13}, those with small velocities, 
should be categorized into the slower component in the two-wave paradigm, where
the faster component in these events was too weak to be detected. At the same
time, a few events in \citet{muhr14} might be fast-mode waves. It should be
noted here that other two factors may also contribute to the velocity 
discrepancy between \citet{nitt13} and \citet{muhr14}: (1) The cadence of
{\em STEREO}/EUVI is much lower, which would underestimate the propagation
velocity \citep{long08}; (2) Muhr et al.'s sample was obtained across the 
solar minimum when the coronal magnetic field is relatively weaker, whereas 
Nitta  et al.'s sample was across the solar maximum when the magnetic field
is stronger.

From these examples, the readers can see how divergent the current
terminologies are. \citet{chen12} proposed to use ``EIT waves" for the slower
component, and ``coronal Moreton waves" for the faster component in the 
two-wave paradigm. Alternatively, we might use ``type I EUV waves" for the 
slower component, and ``type II EUV waves" for the faster component. Readers 
are not obliged to accept such terms, but we do think that it is crucial to 
have different terms for the two types of EUV waves. Otherwise, controversy
will remain forever.

If both the faster and the slower waves appear in one eruption event with the 
high-cadence observations such as from {\em SDO}/AIA, it is very easy to
disentangle them. However, similar to the case of CMEs where the three
components often do not show up simultaneously, the two types of EUV waves are
not always present in one event. In the case of only one wave being detected,
it becomes important while difficult to clarify which type it belongs to. Based
on our limited experience, we propose the following rules. (1) If the wave speed
is always subsonic, e.g., below 186 km s$^{-1}$ for the corona with a 
temperature of 1.5 MK, the observed wave must be the slower component. Since the
low corona has a small plasma $\beta$, we can even claim that if the wave speed
is always below $\sim$300 km s$^{-1}$, it has a high probability to be the
slower component. Note
that there are local regions in the corona where the magnetic field strength is
close to zero, e.g., the regions near magnetic null points, where the fast-mode
wave speed approaches the sound speed, i.e., 186 km s$^{-1}$. But this happens
locally. In most volume of the low corona, the plasma $\beta$ is much less than
unity, otherwise the widely accepted force-free assumption in the coronal 
magnetic extrapolations \citep{low15} is not valid at all; (2) If the wave 
speed is above $\sim$500 km s$^{-1}$, generally the observed wave should be the
 faster MHD wave or shock wave, i.e., the faster component in the two-wave
paradigm; (3) If the wave speed is between the two limits mentioned above, say, 
300--500 km s$^{-1}$, a range where the two types of EUV waves overlap in 
velocity, the situation becomes subtle. It depends on other kinematic 
characteristics, e.g., (i) the slower type of EUV waves would stop near 
magnetic separatrices \citep{dela99, chen06b, chenwu11, whit13}; (ii) The brightest
part of the slower type of EUV waves would rotate anti-clockwise/clockwise in
the active region with negative/positive helicity \citep{attr07}; (iii) The
faster type of EUV waves might become concave outward on the solar disk during
propagation, resulting from wave refraction \citep{xue13}.

\section{What are the drivers of the two types of EUV waves?}

When Moreton waves were discovered in 1960s \citep{more60}, people knew nothing
about CMEs since the phenomenon was discovered in 1970s \citep{tous73}.
Therefore, solar
flares were then ascribed to be the cause of Moreton waves \citep{rams66}. For
the same reason, in Uchida's model, Moreton waves were thought to be blast 
waves generated by the pressure pulse in solar flares \citep{uchi68}. Therefore,
Moreton waves were frequently called flare waves \citep{ziri67, vrsn02}. 
However, whereas some Moreton waves are accompanied by C-class flares 
\citep[e.g.,][]{zhan11}, it is easy to find X-class flares not associated
with Moreton waves, e.g., the 2005 January 14 flare studied by \citet{chen06a}.
In this case, no CME is accompanied. Therefore, after CMEs were widely studied,
it was realized that Moreton waves should be related to CMEs \citep{cliv99, 
chen02}. Strictly speaking, it is the erupting flux rope that generates the 
coronal counterpart of an Moreton wave, i.e., a piston-driven shock wave, which straddles over
the frontal loop of the CME.

Probably influenced by the early explanation for Moreton waves, after ``EIT
waves" were discovered, it was initially thought that they might be generated 
by the pressure pulse of the associated solar flare \citep[e.g.,][]{warm05}, 
although it was already argued that CME as the driver cannot be ruled out
immediately after ``EIT waves" were discovered \citep{thom99, wang00}. In
particular, \citet{dela00} suggested that ``EIT waves" are more related to the
magnetic field evolution during CMEs rather than being driven by solar flares.
One important reason is that more than half ``EIT waves" were associated with
tiny flares, such as A- or B-class flares \citep{cliv05}. It is hard to believe
that these tiny flares can generate global coronal waves. In order to
find out the driver of ``EIT waves", \citet{chen06a} did a test by selecting
14 M- or X-class flares that were not associated with CMEs during solar minima.
These flares are $\sim$1000 times stronger than the A- or B-class flares 
in the 1--8 \AA\ soft X-ray flux. If the pressure pulse in
A- or B-class flares can generate global coronal waves, there is no reason for
M- or X-class flares not to be able to generate global coronal waves. In order
to make the test more convincing, \citet{chen06a} chose those flares during
solar minima since the relatively quieter corona in the background of the
eruption favors the detection of ``EIT waves". \citet{chen06a} found that none
of the 14 strong flares was associated with any ``EIT wave". Therefore, they 
concluded that ``EIT waves" are associated with CMEs. Caution should be taken 
here: people later may occasionally find an ``EIT wave" event that is not 
associated with a visible CME, e.g., the ``EIT wave" event studied by 
\citet{chenwu11}. This is presumably due to the fact that some CMEs are too 
faint and they are missed by coronagraph observations \citep{cheng05, gopal10, 
wang11}.

Furthermore, after comparing the spatial relationship between an ``EIT wave" and
the corresponding CME, \citet{chen09b} found that ``EIT waves" are actually 
cospatial with the CME frontal loop. This was confirmed by \citet{ma09} and
\citet{dai10}. It means that, when viewed from the side, e.g., for an ``EIT 
wave" happening above the solar limb, it should have a domelike
structure as shown by \citet{chen09b} and \citet{vero10}; when an ``EIT wave" is
observed from the top, e.g., for an ``EIT wave" happening on the solar disk, we
can see a ringlike front due to projection effects.

In other models mentioned in Section \ref{model}, it is also argued that ``EIT
waves" are strongly related to CMEs. For example, in the successive reconnection
model \citep{attr07}, ``EIT waves" are formed as the erupting magnetic field
reconnects with the neighboring magnetic loops; in the current shell model
\citep{dela08}, ``EIT waves" are formed at the interface between the erupting
core field and the background envelope field.

\section{Significance of the research on the two types of EUV waves}

We are often confronted with the question: what is the significance to study EUV
waves?

First, they are interesting large-scale phenomena in the solar corona, worthy
to be understood. Second, EUV waves are strongly correlated with CMEs. 
Regarding the slower component of the two EUV waves, for example, \citet{muhr14}
found that 95\% of the ``EIT waves" are associated with CMEs. Considering that 
some CMEs, especially the halo events, might be missed by coronagraphs 
\citep{cheng05, lara06, gopal10, wang11}, such a result supports the conclusion
that there is an unambiguous correlation between EIT waves and CMEs 
\citep{bies02}. Regarding the faster component of the two EUV waves, CMEs are 
always associated with the LCPFs studied by \citet{nitt13}. Therefore, either 
the faster or the slower EUV waves, along with the associated large-scale 
dimmings, are perfect indicators of CMEs, especially those coming from the 
front side of the Sun, which have a high chance to hit the Earth. Therefore, 
in case no coronagraphs are working routinely, EUV imagers can provide reliable
information on the possible CMEs from the visible side of the solar disk, which
is crucial for space weather forecasting. Third, CMEs are often observed well
above the solar limb with coronagraphs occulting the solar disk, the initiation
process, which is an extremely important stage in the CME evolution, is often 
missing in CME observations. Fortunately, it was proposed that the slower 
component of the EUV waves, i.e., the diffuse ``EIT wave", is actually the EUV 
counterpart of the CME frontal loop \citep{chen09b}. Therefore, the 
observations of the diffuse ``EIT waves" can fill in the gap, and may provide 
crucial information for the studies on CME triggering and acceleration. On the 
other hand, the understanding of the ``EIT waves" can shed light on our 
understanding of the nature of the CME frontal loop. For example, 
\citet{chen09b} proposed that in the low corona, say, below a heliocentric 
distance of 2.5$R_\odot$, the propagation of the CME frontal loop might be an 
apparent motion, and its radial velocity in the coronagraph images is not the 
real plasma velocity. The real plasma velocity might be $\sim$3 times smaller 
than the apparent velocity often registered in all CME catalogs. Fourth, like 
any other wave phenomenon, EUV waves can also be utilized to diagnose the
parameters of the corona, e.g., the magnetic field, which cannot not be 
measured directly, and some sub-resolution density structures related to 
the evolution of the amplitude and width of EUV fronts \citep{yuan15}.
 This is called coronal seismology. Compared to the local
seismology in which waves trapped in coronal loops are utilized \citep{naka05},
the coronal seismology using EUV waves has the advantage of the large scales 
since they are observed to cover a major part of the solar disk. However, 
caution has to be taken when doing the global coronal seismology using EUV 
waves, since the prerequisite in doing coronal seismology is that we have to be
sure what mode the wave is. Whereas there is no objection on the fast-mode wave
nature for the faster type of EUV waves, colleagues still have different 
opinions on the nature of the slower type of EUV waves. When one uses the 
faster type of EUV waves, with the assumption of a weak fast-mode shock wave, 
one can derive the reasonable magnetic field strength in the solar corona as 
done by \citet{uchi68}. For example, \citet{long13} applied this type of coronal
seismology to a fast-moving ($v=658$ km s$^{-1}$) EUV wave and obtained the
quiet coronal magnetic field strength in the range 2--6 G. Note that they used
the term ``EIT wave" for the wave phenomenon studied in their paper, but we
personally tend to think that it is a faster type of EUV wave. If one applies
the same fast-mode wave assumption to the slower diffuse ``EIT waves", one may
greatly underestimate the coronal magnetic field. These might include 
\citet{ball08} and \citet{west11} among others.

Doing coronal seismology with the slower diffuse ``EIT waves" is not
impossible, but it is really a formidable task, never so straightforward as the
faster type of EUV waves. It strongly depends on our understanding of the
nature of this wavelike apparent phenomenon. So far, several models have been
proposed to explain the slower type of EUV waves. It is an interesting issue
how these models can be utilized in coronal seismology. One effort was tried by
\citet{chen05b} in the framework of the magnetic fieldline stretching model. 
Even for the magnetic fieldline stretching model, the inversion from the 
observed ``EIT wave" velocity to the magnetic field strength is extremely 
difficult. Considering the difficulty of such inversion, \citet{chen09a} 
performed a forward modeling, and derive the velocity profile for the slower
type of EUV wave for a given 2-dimensional magnetic distribution. In principle,
we might adjust the coronal magnetic field in order to make the derived 
velocity of the slower type of EUV wave best match the observation, and then
consider the corresponding 3-dimensional magnetic field to represent the real 
corona.

\section{Summary}

To summarize, here we list some important points of this review paper:

(1) As a CME erupts, at least two types of wavelike phenomena might be
discernable in EUV images. One is the piston-driven shock wave with a velocity
in the range of several hundred to more than 1000 km s$^{-1}$, the other
is not an MHD wave, and the speed is generally below 500 km s$^{-1}$, sometimes
even down to $\sim$10 km s$^{-1}$.
 
(2) Whereas the faster EUV is well believed to be a fast-mode MHD shock wave, 
the nature of the slower wave is still controversial. Even if people agree that
the slower EUV wave is actually the CME frontal loop, the formation of the CME 
frontal loop is still a question to be addressed.  We tend to believe that
the slower EUV wave is an apparent motion, due mainly to successive 
magnetic fieldline stretching, and sometimes to magnetic reconnection.

(3) We feel that the current situation is confusing, and the confusion results
 from the casual usage of the terminologies for EUV waves. We propose that in
 order to avoid further confusion we need two different names for the two
 types of EUV waves. One simple option is that they are classified as slow EUV
 waves and fast EUV waves. The second option is that the slower EUV waves can
 be called ``type I EUV waves" and the faster EUV waves be called ``type II
 EUV waves" since the slower EUV waves should be related to type I radio
 bursts (both of them result from magnetic field reconfiguration) and the
 faster EUV waves should be related to type II radio bursts (both of them are
 from the CME-driven shock, although one may appear without the other, Nitta
 et al. \citeyear{nitt14}). The third
 option is that the slower EUV waves can be called ``coronal propagating
 fronts" and the faster EUV waves can be called fast-mode EUV wave/shock wave.

\begin{acknowledgments}
The author thanks the organizers for the invitation and two anonymous referees
for their thoughtful comments. This research was supported by the grants NSFC 
(11533005, 11025314, and 10933003) and Jiangsu 333 Project.
\end{acknowledgments}

\bibliographystyle{agu08notitle2}
\bibliography{ref}

\begin{thebibliography}{121}
\providecommand{\natexlab}[1]{#1}
\expandafter\ifx\csname urlstyle\endcsname\relax
  \providecommand{\doi}[1]{doi:\discretionary{}{}{}#1}\else
  \providecommand{\doi}{doi:\discretionary{}{}{}\begingroup
  \urlstyle{rm}\Url}\fi

\bibitem[{\textit{{Asai} et~al.}(2012)}]{asai12}
{Asai}, A., et~al. (2012), \textit{\apjl}, \textit{745}, L18,
  \doi{10.1088/2041-8205/745/2/L18}.

\bibitem[{\textit{{Aschwanden} et~al.}(1999)}]{asch99}
{Aschwanden}, M.~J., et~al. (1999), \textit{\apj}, \textit{520}, 880--894,
  \doi{10.1086/307502}.

\bibitem[{\textit{{Attrill} et~al.}(2007)}]{attr07}
{Attrill}, G.~D.~R., et~al. (2007), \textit{Astronomische Nachrichten},
  \textit{328}, 760, \doi{10.1002/asna.200710794}.

\bibitem[{\textit{{Attrill} et~al.}(2014)}]{attr14}
{Attrill}, G.~D.~R., et~al. (2014), \textit{\apj}, \textit{796}, 55,
  \doi{10.1088/0004-637X/796/1/55}.

\bibitem[{\textit{{Balasubramaniam} et~al.}(2007)}]{bala07}
{Balasubramaniam}, K.~S., et~al. (2007), \textit{\apj}, \textit{658},
  1372--1379, \doi{10.1086/512001}.

\bibitem[{\textit{{Ballai} and {Douglas}}(2008)}]{ball08}
{Ballai}, I., and M.~{Douglas} (2008), in \textit{IAU Symposium}, \textit{IAU
  Symposium}, vol. 247, edited by R.~{Erd{\'e}lyi} and C.~A. {Mendoza-Briceno},
  pp. 243--250, \doi{10.1017/S1743921308014932}.

\bibitem[{\textit{Banerjee and Krishna~Prasad}(2015)}]{bane15}
Banerjee, D., and S.~Krishna~Prasad (2015), \textit{This volume}, A04104.

\bibitem[{\textit{{Biesecker} et~al.}(2002)}]{bies02}
{Biesecker}, D.~A., et~al. (2002), \textit{\apj}, \textit{569}, 1009--1015,
  \doi{10.1086/339402}.

\bibitem[{\textit{{Carmichael}}(1964)}]{carm64}
{Carmichael}, H. (1964), \textit{NASA Special Publication}, \textit{50}, 451.

\bibitem[{\textit{{Chen} et~al.}(2010)}]{chenf10}
{Chen}, F., et~al. (2010), \textit{\apj}, \textit{720}, 1254--1261,
  \doi{10.1088/0004-637X/720/2/1254}.

\bibitem[{\textit{{Chen} et~al.}(2011)}]{chenf11}
{Chen}, F., et~al. (2011), \textit{\apj}, \textit{740}, 116,
  \doi{10.1088/0004-637X/740/2/116}.

\bibitem[{\textit{{Chen}}(2006)}]{chen06a}
{Chen}, P.~F. (2006), \textit{\apjl}, \textit{641}, L153--L156,
  \doi{10.1086/503868}.

\bibitem[{\textit{{Chen}}(2009{\natexlab{a}})}]{chen09b}
{Chen}, P.~F. (2009{\natexlab{a}}), \textit{\apjl}, \textit{698}, L112--L115,
  \doi{10.1088/0004-637X/698/2/L112}.

\bibitem[{\textit{{Chen}}(2009{\natexlab{b}})}]{chen09a}
{Chen}, P.~F. (2009{\natexlab{b}}), \textit{Science in China: Physics,
  Mechanics and Astronomy}, \textit{52}, 1785--1789,
  \doi{10.1007/s11433-009-0240-9}.

\bibitem[{\textit{{Chen}}(2011)}]{chen11}
{Chen}, P.~F. (2011), \textit{Living Reviews in Solar Physics}, \textit{8}, 1,
  \doi{10.12942/lrsp-2011-1}.

\bibitem[{\textit{{Chen} and {Fang}}(2012)}]{chen12}
{Chen}, P.~F., and C.~{Fang} (2012), in \textit{EAS Publications Series},
  \textit{EAS Publications Series}, vol.~55, edited by M.~{Faurobert},
  C.~{Fang}, and T.~{Corbard}, pp. 313--320, \doi{10.1051/eas/1255043}.

\bibitem[{\textit{{Chen} and {Shibata}}(2002)}]{chensh02}
{Chen}, P.~F., and K.~{Shibata} (2002), in \textit{8th Asian-Pacific Regional
  Meeting, Volume II}, edited by S.~{Ikeuchi}, J.~{Hearnshaw}, and T.~{Hanawa},
  pp. 421--422.

\bibitem[{\textit{{Chen} and {Wu}}(2011)}]{chenwu11}
{Chen}, P.~F., and Y.~{Wu} (2011), \textit{\apjl}, \textit{732}, L20,
  \doi{10.1088/2041-8205/732/2/L20}.

\bibitem[{\textit{{Chen} et~al.}(2002)}]{chen02}
{Chen}, P.~F., et~al. (2002), \textit{\apjl}, \textit{572}, L99--L102,
  \doi{10.1086/341486}.

\bibitem[{\textit{{Chen} et~al.}(2005{\natexlab{a}})}]{chen05a}
{Chen}, P.~F., et~al. (2005{\natexlab{a}}), \textit{\ssr}, \textit{121},
  201--211, \doi{10.1007/s11214-006-3911-0}.

\bibitem[{\textit{{Chen} et~al.}(2005{\natexlab{b}})}]{chen05b}
{Chen}, P.~F., et~al. (2005{\natexlab{b}}), \textit{\apj}, \textit{622},
  1202--1210, \doi{10.1086/428084}.

\bibitem[{\textit{{Chen} et~al.}(2006)}]{chen06b}
{Chen}, P.~F., et~al. (2006), \textit{Advances in Space Research}, \textit{38},
  456--460, \doi{10.1016/j.asr.2005.01.049}.

\bibitem[{\textit{{Cheng} et~al.}(2005)}]{cheng05}
{Cheng}, J.~X., et~al. (2005), in \textit{Coronal and Stellar Mass Ejections},
  \textit{IAU Symposium}, vol. 226, edited by K.~{Dere}, J.~{Wang}, and
  Y.~{Yan}, pp. 112--113, \doi{10.1017/S1743921305000244}.

\bibitem[{\textit{{Cheng} et~al.}(2012)}]{cheng12}
{Cheng}, X., et~al. (2012), \textit{\apjl}, \textit{745}, L5,
  \doi{10.1088/2041-8205/745/1/L5}.

\bibitem[{\textit{{Cliver} et~al.}(1999)}]{cliv99}
{Cliver}, E.~W., et~al. (1999), \textit{\solphys}, \textit{187}, 89--114,
  \doi{10.1023/A:1005115119661}.

\bibitem[{\textit{{Cliver} et~al.}(2005)}]{cliv05}
{Cliver}, E.~W., et~al. (2005), \textit{\apj}, \textit{631}, 604--611,
  \doi{10.1086/432250}.

\bibitem[{\textit{{Cohen} et~al.}(2009)}]{cohe09}
{Cohen}, O., et~al. (2009), \textit{\apj}, \textit{705}, 587,
  \doi{10.1088/0004-637X/705/1/587}.

\bibitem[{\textit{{Dai} et~al.}(2010)}]{dai10}
{Dai}, Y., et~al. (2010), \textit{\apj}, \textit{708}, 913--919,
  \doi{10.1088/0004-637X/708/2/913}.

\bibitem[{\textit{{Delaboudini{\`e}re} et~al.}(1995)}]{dela95}
{Delaboudini{\`e}re}, J.-P., et~al. (1995), \textit{\solphys}, \textit{162},
  291--312, \doi{10.1007/BF00733432}.

\bibitem[{\textit{{Delann{\'e}e}}(2000)}]{dela00}
{Delann{\'e}e}, C. (2000), \textit{\apj}, \textit{545}, 512--523,
  \doi{10.1086/317777}.

\bibitem[{\textit{{Delann{\'e}e} and {Aulanier}}(1999)}]{dela99}
{Delann{\'e}e}, C., and G.~{Aulanier} (1999), \textit{\solphys}, \textit{190},
  107--129, \doi{10.1023/A:1005249416605}.

\bibitem[{\textit{{Delann{\'e}e} et~al.}(2008)}]{dela08}
{Delann{\'e}e}, C., et~al. (2008), \textit{\solphys}, \textit{247}, 123--150,
  \doi{10.1007/s11207-007-9085-4}.

\bibitem[{\textit{{Downs} et~al.}(2012)}]{down12}
{Downs}, C., et~al. (2012), \textit{\apj}, \textit{750}, 134,
  \doi{10.1088/0004-637X/750/2/134}.

\bibitem[{\textit{{Eto} et~al.}(2002)}]{eto02}
{Eto}, S., et~al. (2002), \textit{\pasj}, \textit{54}, 481--491,
  \doi{10.1093/pasj/54.3.481}.

\bibitem[{\textit{{Fang} et~al.}(2013)}]{fang13}
{Fang}, C., et~al. (2013), \textit{Research in Astronomy and Astrophysics},
  \textit{13}, 1509, \doi{10.1088/1674-4527/13/12/011}.

\bibitem[{\textit{{Forbes}}(2000)}]{forb00}
{Forbes}, T.~G. (2000), \textit{\jgr}, \textit{105}, 23,153--23,166,
  \doi{10.1029/2000JA000005}.

\bibitem[{\textit{{Francile} et~al.}(2013)}]{fran13}
{Francile}, C., et~al. (2013), \textit{\aap}, \textit{552}, A3,
  \doi{10.1051/0004-6361/201118001}.

\bibitem[{\textit{{Gallagher} and {Long}}(2011)}]{gall11}
{Gallagher}, P.~T., and D.~M. {Long} (2011), \textit{\ssr}, \textit{158},
  365--396, \doi{10.1007/s11214-010-9710-7}.

\bibitem[{\textit{Gilbert et~al.}(2008)}]{gilb08}
Gilbert, H.~R., et~al. (2008), \textit{The Astrophysical Journal},
  \textit{685}(1), 629.

\bibitem[{\textit{{Gopalswamy} et~al.}(2009)}]{gopal09}
{Gopalswamy}, N., et~al. (2009), \textit{\apjl}, \textit{691}, L123--L127,
  \doi{10.1088/0004-637X/691/2/L123}.

\bibitem[{\textit{{Gopalswamy} et~al.}(2010)}]{gopal10}
{Gopalswamy}, N., et~al. (2010), \textit{Twelfth International Solar Wind
  Conference}, \textit{1216}, 452--458, \doi{10.1063/1.3395902}.

\bibitem[{\textit{{Grechnev} et~al.}(2011)}]{grec11}
{Grechnev}, V.~V., et~al. (2011), \textit{\solphys}, \textit{273}, 461--477,
  \doi{10.1007/s11207-011-9781-y}.

\bibitem[{\textit{{Harra} and {Sterling}}(2003)}]{harr03}
{Harra}, L.~K., and A.~C. {Sterling} (2003), \textit{\apj}, \textit{587},
  429--438, \doi{10.1086/368079}.

\bibitem[{\textit{{Hudson}}(1999)}]{huds99}
{Hudson}, H. (1999), \textit{\solphys}, \textit{190}, 91--106,
  \doi{10.1023/A:1005246501003}.

\bibitem[{\textit{{Klassen} et~al.}(2000)}]{klas00}
{Klassen}, A., et~al. (2000), \textit{\aaps}, \textit{141}, 357--369,
  \doi{10.1051/aas:2000125}.

\bibitem[{\textit{{Krucker} et~al.}(1999)}]{kruc99}
{Krucker}, S., et~al. (1999), \textit{\apj}, \textit{519}, 864--875,
  \doi{10.1086/307415}.

\bibitem[{\textit{{Kumar} et~al.}(2013)}]{kuma13}
{Kumar}, P., et~al. (2013), \textit{\solphys}, \textit{282}, 523--541,
  \doi{10.1007/s11207-012-0158-7}.

\bibitem[{\textit{{Kwon} et~al.}(2013)}]{kwon13}
{Kwon}, R.-Y., et~al. (2013), \textit{\apj}, \textit{766}, 55,
  \doi{10.1088/0004-637X/766/1/55}.

\bibitem[{\textit{{Lara} et~al.}(2006)}]{lara06}
{Lara}, A., et~al. (2006), \textit{Journal of Geophysical Research (Space
  Physics)}, \textit{111}, A06107, \doi{10.1029/2005JA011431}.

\bibitem[{\textit{{Li} et~al.}(2012)}]{li12}
{Li}, T., et~al. (2012), \textit{\apj}, \textit{746}, 13,
  \doi{10.1088/0004-637X/746/1/13}.

\bibitem[{\textit{{Liu} and {Ofman}}(2014)}]{liu14}
{Liu}, W., and L.~{Ofman} (2014), \textit{\solphys}, \textit{289}, 3233--3277,
  \doi{10.1007/s11207-014-0528-4}.

\bibitem[{\textit{{Liu} et~al.}(2010)}]{liu10}
{Liu}, W., et~al. (2010), \textit{\apjl}, \textit{723}, L53--L59,
  \doi{10.1088/2041-8205/723/1/L53}.

\bibitem[{\textit{{Liu} et~al.}(2012)}]{liu12}
{Liu}, W., et~al. (2012), \textit{\apj}, \textit{753}, 52,
  \doi{10.1088/0004-637X/753/1/52}.

\bibitem[{\textit{{Long} et~al.}(2008)}]{long08}
{Long}, D.~M., et~al. (2008), \textit{\apjl}, \textit{680}, L81--L84,
  \doi{10.1086/589742}.

\bibitem[{\textit{{Long} et~al.}(2013)}]{long13}
{Long}, D.~M., et~al. (2013), \textit{\solphys}, \textit{288}, 567--583,
  \doi{10.1007/s11207-013-0331-7}.

\bibitem[{\textit{{Low}}(2015)}]{low15}
{Low}, B.~C. (2015), \textit{Science China Physics, Mechanics, and Astronomy},
  \textit{58}, 2, \doi{10.1007/s11433-014-5626-7}.

\bibitem[{\textit{{Ma} et~al.}(2009)}]{ma09}
{Ma}, S., et~al. (2009), \textit{\apj}, \textit{707}, 503--509,
  \doi{10.1088/0004-637X/707/1/503}.

\bibitem[{\textit{{Meyer}}(1968)}]{meye68}
{Meyer}, F. (1968), in \textit{Structure and Development of Solar Active
  Regions}, \textit{IAU Symposium}, vol.~35, edited by K.~O. {Kiepenheuer}, p.
  485.

\bibitem[{\textit{{Moreton} and {Ramsey}}(1960)}]{more60}
{Moreton}, G.~E., and H.~E. {Ramsey} (1960), \textit{\pasp}, \textit{72}, 357,
  \doi{10.1086/127549}.

\bibitem[{\textit{{Moses} et~al.}(1997)}]{mose97}
{Moses}, D., et~al. (1997), \textit{\solphys}, \textit{175}, 571--599,
  \doi{10.1023/A:1004902913117}.

\bibitem[{\textit{{Muhr} et~al.}(2014)}]{muhr14}
{Muhr}, N., et~al. (2014), \textit{\solphys}, \textit{289}, 4563--4588,
  \doi{10.1007/s11207-014-0594-7}.

\bibitem[{\textit{{Nakariakov} and {Verwichte}}(2005)}]{naka05}
{Nakariakov}, V.~M., and E.~{Verwichte} (2005), \textit{Living Reviews in Solar
  Physics}, \textit{2}, 3, \doi{10.12942/lrsp-2005-3}.

\bibitem[{\textit{{Nakariakov} et~al.}(1999)}]{naka99}
{Nakariakov}, V.~M., et~al. (1999), \textit{Science}, \textit{285}, 862--864,
  \doi{10.1126/science.285.5429.862}.

\bibitem[{\textit{{Nitta} et~al.}(2013)}]{nitt13}
{Nitta}, N.~V., et~al. (2013), \textit{\apj}, \textit{776}, 58,
  \doi{10.1088/0004-637X/776/1/58}.

\bibitem[{\textit{{Nitta} et~al.}(2014)}]{nitt14}
{Nitta}, N.~V., et~al. (2014), \textit{\solphys}, \textit{289}, 4589--4606,
  \doi{10.1007/s11207-014-0602-y}.

\bibitem[{\textit{{Ofman} and {Thompson}}(2002)}]{ofma02}
{Ofman}, L., and B.~J. {Thompson} (2002), \textit{\apj}, \textit{574},
  440--452, \doi{10.1086/340924}.

\bibitem[{\textit{{Patsourakos} and {Vourlidas}}(2009)}]{pats09a}
{Patsourakos}, S., and A.~{Vourlidas} (2009), \textit{\apjl}, \textit{700},
  L182--L186, \doi{10.1088/0004-637X/700/2/L182}.

\bibitem[{\textit{{Patsourakos} and {Vourlidas}}(2012)}]{pats12}
{Patsourakos}, S., and A.~{Vourlidas} (2012), \textit{\solphys}, \textit{281},
  187--222, \doi{10.1007/s11207-012-9988-6}.

\bibitem[{\textit{{Patsourakos} et~al.}(2009)}]{pats09b}
{Patsourakos}, S., et~al. (2009), \textit{\solphys}, \textit{259}, 49--71,
  \doi{10.1007/s11207-009-9386-x}.

\bibitem[{\textit{{Patsourakos} et~al.}(2010)}]{pats10}
{Patsourakos}, S., et~al. (2010), \textit{\apjl}, \textit{724}, L188--L193,
  \doi{10.1088/2041-8205/724/2/L188}.

\bibitem[{\textit{{Podladchikova} and {Berghmans}}(2005)}]{podl05}
{Podladchikova}, O., and D.~{Berghmans} (2005), \textit{\solphys},
  \textit{228}, 265--284, \doi{10.1007/s11207-005-5373-z}.

\bibitem[{\textit{{Pohjolainen} et~al.}(2001)}]{pohj01}
{Pohjolainen}, S., et~al. (2001), \textit{\apj}, \textit{556}, 421--431,
  \doi{10.1086/321577}.

\bibitem[{\textit{{Ramsey} and {Smith}}(1966)}]{rams66}
{Ramsey}, H.~E., and S.~F. {Smith} (1966), \textit{\aj}, \textit{71}, 197,
  \doi{10.1086/109903}.

\bibitem[{\textit{{Schrijver} et~al.}(2011)}]{schr11}
{Schrijver}, C.~J., et~al. (2011), \textit{\apj}, \textit{738}, 167,
  \doi{10.1088/0004-637X/738/2/167}.

\bibitem[{\textit{{Selwa} et~al.}(2013)}]{selw13}
{Selwa}, M., et~al. (2013), \textit{\solphys}, \textit{284}, 515--539,
  \doi{10.1007/s11207-013-0302-z}.

\bibitem[{\textit{{Shen} and {Liu}}(2012)}]{shen12}
{Shen}, Y., and Y.~{Liu} (2012), \textit{\apjl}, \textit{752}, L23,
  \doi{10.1088/2041-8205/752/2/L23}.

\bibitem[{\textit{{Shen} et~al.}(2013)}]{shen13}
{Shen}, Y., et~al. (2013), \textit{\apjl}, \textit{773}, L33,
  \doi{10.1088/2041-8205/773/2/L33}.

\bibitem[{\textit{{Sterling} and {Hudson}}(1997)}]{ster97}
{Sterling}, A.~C., and H.~S. {Hudson} (1997), \textit{\apjl}, \textit{491},
  L55--L58, \doi{10.1086/311043}.

\bibitem[{\textit{{Svestka}}(1976)}]{sves76}
{Svestka}, Z. (1976), \textit{Solar Flares}, 1 ed., Springer-Verlag, Berlin
  Heidelberg, an optional note.

\bibitem[{\textit{{Thompson} and {Myers}}(2009)}]{thom09}
{Thompson}, B.~J., and D.~C. {Myers} (2009), \textit{\apjs}, \textit{183},
  225--243, \doi{10.1088/0067-0049/183/2/225}.

\bibitem[{\textit{{Thompson} et~al.}(1998)}]{thom98}
{Thompson}, B.~J., et~al. (1998), \textit{\grl}, \textit{25}, 2465--2468,
  \doi{10.1029/98GL50429}.

\bibitem[{\textit{{Thompson} et~al.}(1999)}]{thom99}
{Thompson}, B.~J., et~al. (1999), \textit{\apjl}, \textit{517}, L151--L154,
  \doi{10.1086/312030}.

\bibitem[{\textit{{Thompson} et~al.}(2000)}]{thom00}
{Thompson}, B.~J., et~al. (2000), \textit{\solphys}, \textit{193}, 161--180,
  \doi{10.1023/A:1005222123970}.

\bibitem[{\textit{{Torsti} et~al.}(1999)}]{tors99}
{Torsti}, J., et~al. (1999), \textit{\apj}, \textit{510}, 460--465,
  \doi{10.1086/306581}.

\bibitem[{\textit{{Tousey}}(1973)}]{tous73}
{Tousey}, R. (1973), in \textit{Space Research Conference}, edited by M.~J.
  {Rycroft} and S.~K. {Runcorn}, pp. 713--730.

\bibitem[{\textit{{Tripathi} and {Raouafi}}(2007)}]{trip07}
{Tripathi}, D., and N.-E. {Raouafi} (2007), \textit{\aap}, \textit{473},
  951--957, \doi{10.1051/0004-6361:20077255}.

\bibitem[{\textit{{Uchida}}(1968)}]{uchi68}
{Uchida}, Y. (1968), \textit{\solphys}, \textit{4}, 30--44,
  \doi{10.1007/BF00146996}.

\bibitem[{\textit{{Uchida}}(1974)}]{uchi74}
{Uchida}, Y. (1974), \textit{\solphys}, \textit{39}, 431--449,
  \doi{10.1007/BF00162436}.

\bibitem[{\textit{{UeNo} et~al.}(2004)}]{ueno04}
{UeNo}, S., et~al. (2004), in \textit{Ground-based Instrumentation for
  Astronomy}, \textit{Society of Photo-Optical Instrumentation Engineers (SPIE)
  Conference Series}, vol. 5492, edited by A.~F.~M. {Moorwood} and M.~{Iye},
  pp. 958--969, \doi{10.1117/12.550304}.

\bibitem[{\textit{{Vernazza} et~al.}(1981)}]{val81}
{Vernazza}, J.~E., et~al. (1981), \textit{\apjs}, \textit{45}, 635--725,
  \doi{10.1086/190731}.

\bibitem[{\textit{{Veronig} et~al.}(2010)}]{vero10}
{Veronig}, A.~M., et~al. (2010), \textit{\apjl}, \textit{716}, L57--L62,
  \doi{10.1088/2041-8205/716/1/L57}.

\bibitem[{\textit{{Vourlidas} et~al.}(2003)}]{vour03}
{Vourlidas}, A., et~al. (2003), \textit{\apj}, \textit{598}, 1392--1402,
  \doi{10.1086/379098}.

\bibitem[{\textit{{Vr{\v s}nak} et~al.}(2002)}]{vrsn02}
{Vr{\v s}nak}, B., et~al. (2002), \textit{\aap}, \textit{394}, 299--310,
  \doi{10.1051/0004-6361:20021121}.

\bibitem[{\textit{{Wang} et~al.}(2009)}]{wang09}
{Wang}, H., et~al. (2009), \textit{\apj}, \textit{700}, 1716--1731,
  \doi{10.1088/0004-637X/700/2/1716}.

\bibitem[{\textit{{Wang}}(2015)}]{wang15}
{Wang}, T. (2015), \textit{This volume}, \textit{116}, A04104.

\bibitem[{\textit{{Wang} et~al.}(2011)}]{wang11}
{Wang}, Y., et~al. (2011), \textit{Journal of Geophysical Research (Space
  Physics)}, \textit{116}, A04104, \doi{10.1029/2010JA016101}.

\bibitem[{\textit{{Wang}}(2000)}]{wang00}
{Wang}, Y.-M. (2000), \textit{\apjl}, \textit{543}, L89--L93,
  \doi{10.1086/318178}.

\bibitem[{\textit{{Warmuth}}(2010)}]{warm10}
{Warmuth}, A. (2010), \textit{Advances in Space Research}, \textit{45},
  527--536, \doi{10.1016/j.asr.2009.08.022}.

\bibitem[{\textit{{Warmuth} and {Mann}}(2011)}]{warm11}
{Warmuth}, A., and G.~{Mann} (2011), \textit{\aap}, \textit{532}, A151,
  \doi{10.1051/0004-6361/201116685}.

\bibitem[{\textit{{Warmuth} et~al.}(2005)}]{warm05}
{Warmuth}, A., et~al. (2005), \textit{\apjl}, \textit{626}, L121--L124,
  \doi{10.1086/431756}.

\bibitem[{\textit{{Webb}}(2000)}]{webb00}
{Webb}, D.~F. (2000), \textit{Journal of Atmospheric and Solar-Terrestrial
  Physics}, \textit{62}, 1415--1426, \doi{10.1016/S1364-6826(00)00075-4}.

\bibitem[{\textit{{West} et~al.}(2011)}]{west11}
{West}, M.~J., et~al. (2011), \textit{\apj}, \textit{730}, 122,
  \doi{10.1088/0004-637X/730/2/122}.

\bibitem[{\textit{{White} and {Thompson}}(2005)}]{whit05}
{White}, S.~M., and B.~J. {Thompson} (2005), \textit{\apjl}, \textit{620},
  L63--L66, \doi{10.1086/428428}.

\bibitem[{\textit{{White} et~al.}(2013)}]{whit13}
{White}, S.~M., et~al. (2013), \textit{Technical report of Air Force Research
  Laboratory}, \textit{22}, 1--22.

\bibitem[{\textit{{Wild} et~al.}(1963)}]{wild63}
{Wild}, J.~P., et~al. (1963), \textit{\araa}, \textit{1}, 291,
  \doi{10.1146/annurev.aa.01.090163.001451}.

\bibitem[{\textit{{Wills-Davey}}(2006)}]{will06}
{Wills-Davey}, M.~J. (2006), \textit{\apj}, \textit{645}, 757--765,
  \doi{10.1086/504144}.

\bibitem[{\textit{{Wills-Davey} and {Attrill}}(2009)}]{will09}
{Wills-Davey}, M.~J., and G.~D.~R. {Attrill} (2009), \textit{\ssr},
  \textit{149}, 325--353, \doi{10.1007/s11214-009-9612-8}.

\bibitem[{\textit{{Wills-Davey} and {Thompson}}(1999)}]{will99}
{Wills-Davey}, M.~J., and B.~J. {Thompson} (1999), \textit{\solphys},
  \textit{190}, 467--483, \doi{10.1023/A:1005201500675}.

\bibitem[{\textit{{Wills-Davey} et~al.}(2007)}]{will07}
{Wills-Davey}, M.~J., et~al. (2007), \textit{\apj}, \textit{664}, 556--562,
  \doi{10.1086/519013}.

\bibitem[{\textit{{Wu} et~al.}(2001)}]{wu01}
{Wu}, S.~T., et~al. (2001), \textit{\jgr}, \textit{106}, 25,089--25,102,
  \doi{10.1029/2000JA000447}.

\bibitem[{\textit{{Xue} et~al.}(2013)}]{xue13}
{Xue}, Z.~K., et~al. (2013), \textit{\aap}, \textit{556}, A152,
  \doi{10.1051/0004-6361/201220731}.

\bibitem[{\textit{{Yang} and {Chen}}(2010)}]{yang10}
{Yang}, H.~Q., and P.~F. {Chen} (2010), \textit{\solphys}, \textit{266},
  59--69, \doi{10.1007/s11207-010-9595-3}.

\bibitem[{\textit{{Yang} et~al.}(2013)}]{yang13}
{Yang}, L., et~al. (2013), \textit{\apj}, \textit{775}, 39,
  \doi{10.1088/0004-637X/775/1/39}.

\bibitem[{\textit{{Yuan} et~al.}(2015)}]{yuan15}
{Yuan}, D., et~al. (2015), \textit{\apj}, \textit{799}, 221,
  \doi{10.1088/0004-637X/799/2/221}.

\bibitem[{\textit{{Zhang} et~al.}(2011)}]{zhan11}
{Zhang}, Y., et~al. (2011), \textit{\pasj}, \textit{63}, 685--,
  \doi{10.1093/pasj/63.3.685}.

\bibitem[{\textit{{Zhao} et~al.}(2011)}]{zhao11}
{Zhao}, X.~H., et~al. (2011), \textit{\apj}, \textit{742}, 131,
  \doi{10.1088/0004-637X/742/2/131}.

\bibitem[{\textit{{Zheng} et~al.}(2012{\natexlab{a}})}]{zhen12a}
{Zheng}, R., et~al. (2012{\natexlab{a}}), \textit{\apj}, \textit{747}, 67,
  \doi{10.1088/0004-637X/747/1/67}.

\bibitem[{\textit{{Zheng} et~al.}(2012{\natexlab{b}})}]{zhen12b}
{Zheng}, R., et~al. (2012{\natexlab{b}}), \textit{\apjl}, \textit{753}, L29,
  \doi{10.1088/2041-8205/753/2/L29}.

\bibitem[{\textit{{Zhukov}}(2011)}]{zhuk11}
{Zhukov}, A.~N. (2011), \textit{Journal of Atmospheric and Solar-Terrestrial
  Physics}, \textit{73}, 1096--1116, \doi{10.1016/j.jastp.2010.11.030}.

\bibitem[{\textit{{Zhukov} et~al.}(2009)}]{zhuk09}
{Zhukov}, A.~N., et~al. (2009), \textit{\solphys}, \textit{259}, 73--85,
  \doi{10.1007/s11207-009-9375-0}.

\bibitem[{\textit{{Zirin} and {Werner}}(1967)}]{ziri67}
{Zirin}, H., and S.~{Werner} (1967), \textit{\solphys}, \textit{1}, 66--100,
  \doi{10.1007/BF00150304}.

\end{thebibliography}

\end{article}
 
\clearpage


\clearpage

\clearpage

\clearpage

\clearpage

\end{document}